\documentclass[english,keywords,aps,twocolumn]{revtex4-2}

\usepackage{amsmath}
\usepackage[]{graphicx}
\usepackage{color}
\usepackage[utf8]{inputenc}
\usepackage{marginnote}
\usepackage[UKenglish]{babel}
\usepackage{supertabular}
\usepackage{enumitem}

\usepackage{amssymb}        % AMS Math
\usepackage{textcomp}
\begin{document}
\title{Neutron interference from a split-crystal interferometer}
%\shorttitle{Split-crystal interferometer}
\author{Hartmut Lemmel$^{1,2}$}
\email{hartmut.lemmel@tuwien.ac.at}
\author{Michael Jentschel$^{2}$}
\author{Hartmut Abele$^{1}$}
\author{Fabien Lafont$^{2}$}
\author{Bruno Guerard$^{2}$}
\author{Carlo P. Sasso$^{3}$}
\author{Giovanni Mana$^{3}$}
\author{Enrico Massa$^{3}$}
\email{e.massa@inrim.it}

\affiliation{
$^1$Atominstitut, TU Wien, Stadionallee 2, 1020 Vienna, Austria \\
$^2$ILL -- Institut Laue-Langevin, 38000, Grenoble, France\\
$^3$INRIM -- Istituto Nazionale di Ricerca Metrologica, Torino, Italy
}

%\keyword{split-crystal interferometry}
%\keyword{neutron interferometry}

\begin{abstract}
We report the first successful operation of a neutron interferometer having a separate beam-recombining crystal. We achieved this result at the neutron interferometry setup S18 at the ILL in Grenoble by a collaboration between TU Wien, ILL Grenoble and INRIM Torino. While previous interferometers were machined out of a single crystal block, we managed to align two crystals on nanoradian and picometer scales, as required to obtain neutron interference.
As a decisive proof of principle demonstration, it opens the door to a new generation of neutron interferometers and exciting applications.
\end{abstract}

\maketitle

\section{Introduction}%

\hspace{1mm}\\
Neutron interferometry requires splitting and recombining the wave function of every single neutron while maintaining the coherence of the neutron wave packet. For thermal neutrons, a frequent approach is the use of silicon single crystals. Diffraction allows efficiently to fulfill the task of splitting, reflecting and recombining the beams. The path separation is typically in the order of centimeters.

Since its first demonstration in 1974 by Rauch, Treimer and Bonse \cite{rauch1974}, this technique was used in a broad spectrum of fields  \cite{rauch-werner}. Being based on the superposition of matter waves of a single, electrically neutral particle with spin 1/2, neutron interferometry was used for a large number of critical tests of quantum mechanics, from the $4\pi$ symmetry of fermions \cite{rauch1975} to the demonstration of the Quantum Cheshire Cat \cite{denkmayr2014}. It was used to study the interplay with gravity in the famous COW experiment \cite{colella1975}, to measure neutron scattering lengths \cite{Haun2020} and to search for the existence of hypothetical new forces \cite{lemmel2015,Li2016}.

Achieving neutron interference requires the alignment of the involved crystal planes within a few nanoradians. Relative displacements of more than a few picometers must be avoided. Up to now, these problems were solved by machining the entire interferometer out of a monolithic single crystal. This design -- based on the availability of perfect single crystal ingots -- has limitations: The sensitivity of many experiments depends critically on the area enclosed by the two wave paths and therefore by path separation and path length within the interferometer. Another limitation is the size and complexity of test objects that can be inserted into the interferometer. Finally, the sensitivity is largely impacted by the uniformity of the lattice spacing in the diffracting crystals \cite{Heacock2017}. Large single ingot interferometers \cite{Zawisky2009} showed difficulties associated with strain fields within large crystals.

A promising solution to dramatically improve the sensitivity of neutron interferometers would be an interferometer consisting of separated crystals. An attempt to build such a split-crystal interferometer is reported in \cite {Uebbing1991}, but it did not succeed in achieving neutron interference. Neutron interferometry with physically split gratings using cold or very cold neutrons is reported in \cite{zouw2000,pruner2006}. However, the small separation of the beam paths and/or the low count rates of very cold neutron beams did not allow to have competitive sensitivity.

\begin{figure}\centering
\includegraphics[width=0.95\columnwidth]{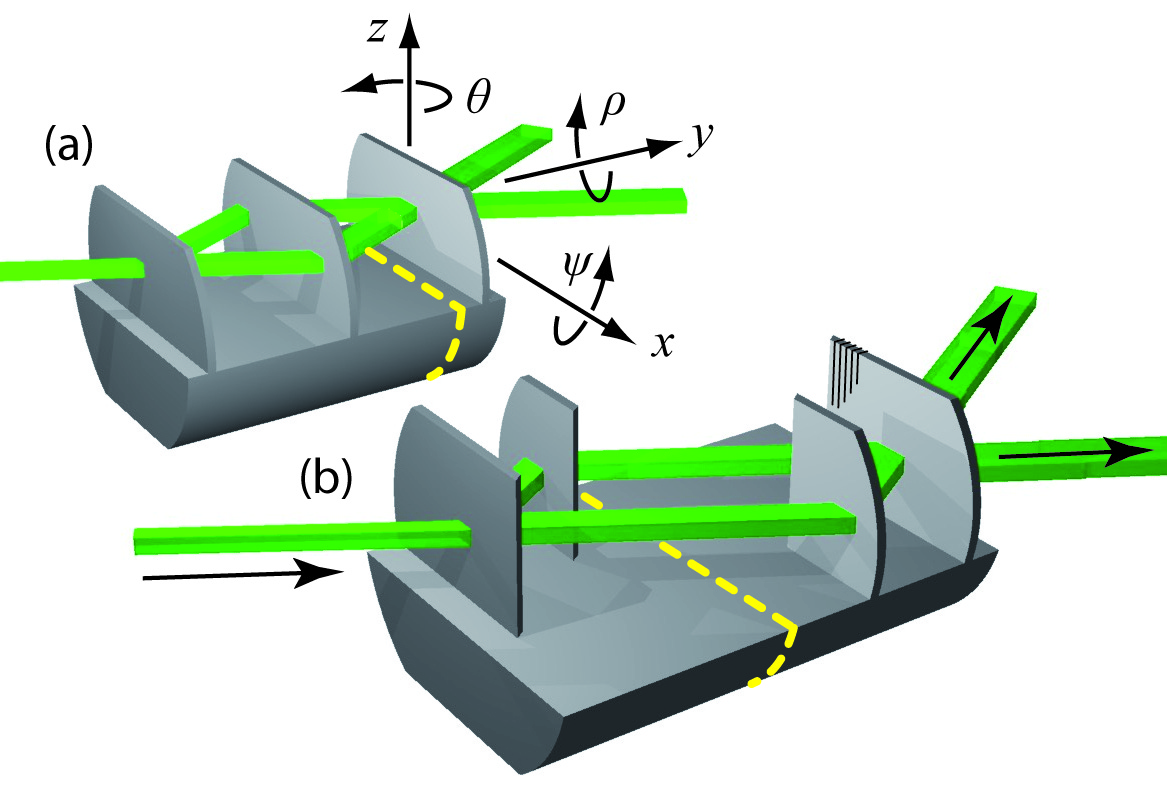}
\caption{Symmetric (a) and skew-symmetric (b) crystal interferometers. The neutrons are coherently split, reflected and recombined by the crystal lamellas. The dashed yellow lines indicate the usual cuts which create a split-crystal interferometer.}\label{figsymm}
\end{figure}

Perfect crystal interferometers are also used with X-rays, firstly demonstrated by Bonse and Hart in 1965 \cite{bonse1965}. Contrary to neutrons, X-rays are substantially absorbed by the crystal. While the loss of intensity is compensated by brighter sources, the absorption creates wider beam acceptance angles. It was therefore possible already in 1968 to create a split-crystal interferometer for X-rays \cite{Bonse1968,deslattes1969}.

In the case of a symmetric interferometer, the third lamella (the analyser) is separated, cf.\ Fig.\ \ref{figsymm} (a). Then the interference signal is extremely sensitive to the analyser's $x$ position, since a movement by one lattice constant ($d \approx 0.192$ nm) creates a phase shift of $2\pi$. Such an interferometer allowed measuring the lattice parameter of $^{28}$Si with parts per billion accuracy \cite{Massa2011,Massa2015} leading to the determination of the Avogadro constant \cite{Fujii2018}, thereby realizing the kilogram by counting atoms \cite{Massa2020}.

Alternatively, two pairs of crystal lamellas can be separated, as shown in Fig.\ \ref{figsymm} (b). These skew-symmetric split-crystals are insensitive to misalignments along the $x$ axis and allow the realisation of long and spaced interferometer arms, as well as scans of the arm length and Bragg angle alignment. They were used for phase-contrast imaging \cite{Yoneyama2002} profiting from the extended sensed area and volume.

Eventually, an interferometer composed of six separate crystals allowed characterizing the temporal coherence of 10 keV pulses from an X-ray free-electron laser \cite{Osaka2017}.

This paper reports on the first successful operation of a split-crystal interferometer using thermal neutrons, showing that all requirements to build such a device are under control. We used an existing symmetric split-crystal interferometer, but the proof of principle demonstration is just as valid for a skew-symmetric setup.

\section{Alignment Requirements}%

\begin{figure}\centering\vspace{1mm}
\includegraphics[width=0.95\columnwidth]{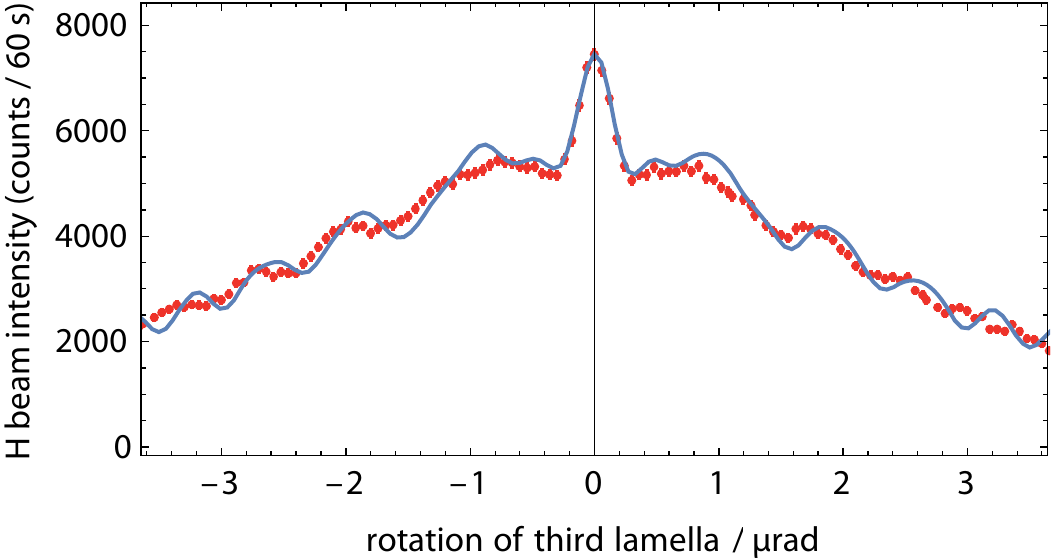}
\caption{Rocking curve of the analyser lamella with path 1 of the interferometer blocked, cf. Fig.\ \ref{figsetup}. The error bars lie within the bullets. The theoretical prediction (solid line) is calculated by convolving the product of three Bragg reflections (the triple bounce Bragg monochromator) and two Laue reflections (splitter and mirror lamella) with the final Laue reflection by the analyser.}\label{figtheta}
\end{figure}

\hspace{1mm}\\
Neutron interference requires that the relative alignment of the two crystals of a symmetric interferometer fulfills the following conditions.
\begin{itemize}[leftmargin=*]
  \item {\it $\theta$, yaw angle, rotation about the vertical axis.} This angle must be adjusted to match the Bragg condition. The observed triple-Laue rocking curve is shown in Fig.\ \ref{figtheta}. The broad peak is modulated by Pendell\"osung fringes which culminate in a central spike \cite{Petrascheck1984}. While X-ray interference is possible within the whole broad peak, neutron interference is only possible in the central spike \cite{mana1997b} which has a full width at half maximum of about 250 nrad.
  \item {\it $\rho$, pitch angle, rotation about the lamella's surface normal.} This angle must be adjusted to make the interfering beams parallel. If misaligned, the beams are slightly inclined to each other and create a horizontal moir\'{e} pattern. To make the interference observable, the fringe spacing $\Lambda_z = d/\rho$ must be greater than the vertical detector resolution, where $d = 0.192$ nm denotes the spacing of the diffracting $\{220\}$ planes of the silicon crystals. The alignment of the $\rho$ angle is not trivial because -- contrary to the $\theta$ angle -- its variation hardly changes the intensity. A smart alignment strategy is described further below.
  \item {\it $\psi$, roll angle, rotation about the diffracting plane normal.} No accurate adjustment is required. The interferometer is insensitive to this misalignment unless it becomes macroscopic.
  \item {\it $x$, axial position along the diffracting plane normal.} Statically, this degree of freedom is unessential but the noise or drift during the measurement time must be less than a few picometers, since a displacement of the analyser by one diffracting plane creates a full interference period. This applies only to a symmetric interferometer. In the skew-symmetric layout, all translational degrees of freedom cancel making it the best choice for a large-scale interferometer.
  \item {\it $y$, transverse position along the lamella's surface normal.} To avoid defocusing, the analyser-to-mirrors distance must be equal to the splitter-to-mirrors one to within a few micrometers. This applies only to a symmetric interferometer. In the skew-symmetric layout, the two crystals can be spaced at will.
  \item {\it $z$, vertical position.} No adjustment is required.
  \item The lattice constant of the two crystals must be equal, otherwise vertical moir\'{e} fringes occur, which are spaced by $d/\epsilon_{xx}$, where $\epsilon_{xx}=\alpha \Delta T$ is the thermal strain between the crystals. Taking into account the thermal expansion of silicon, $\alpha=2.5 \times 10^{-6}$ / K, this means that the temperature difference $\Delta T$ of the two crystals must be less than 10 mK.
\end{itemize}

All these parameters must not only be aligned but also kept constant over a typical measurement time. Alternatively, if a parameter is drifting but can be monitored, a time-dependent neutron detection can be used to reconstruct the correct phase.

\section{Experimental setup}

\begin{figure}\centering
\includegraphics[width=0.95\columnwidth]{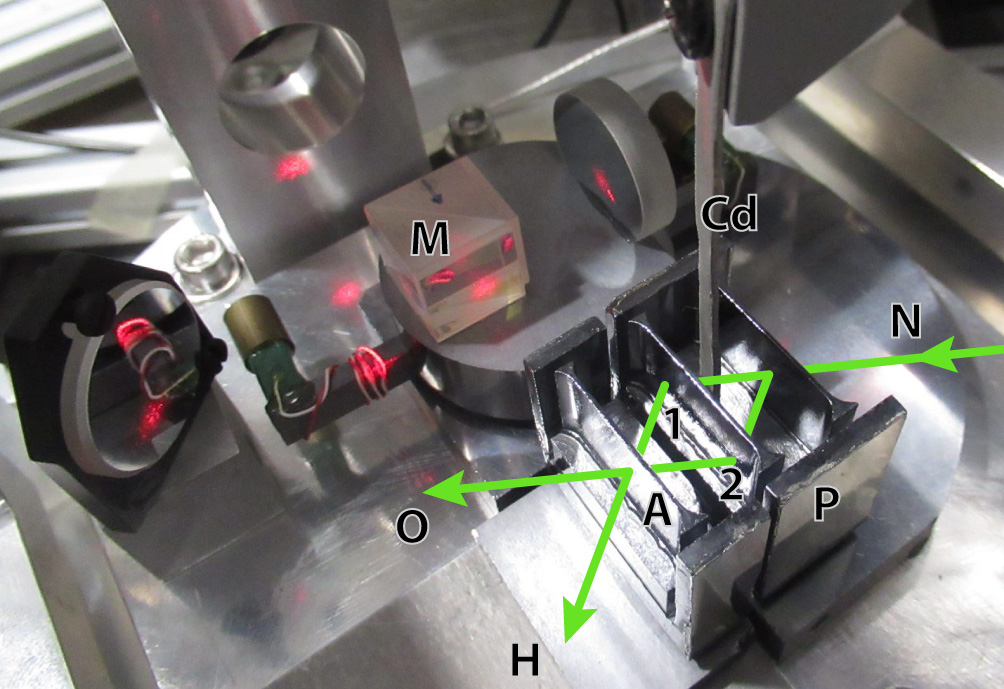}
\caption{The crystal interferometer with separate analyser lamella (A). The neutron beam (N) enters from the right side, is split into two paths (1, 2), and exits through the output ports O and H. Polished surfaces on the crystal's sides allow pre-alignment by an optical autocollimator on the front (P) and monitoring the analyser's coordinates ($\theta$, $\rho$ and $x$) by an optical interferometer (M) on the rear. A piece of cadmium (Cd) can be inserted from the top to block one or the other interferometer path.}\label{figsetup}
\end{figure}

\hspace{1mm}\\
The setup is depicted in Fig.\ \ref{figsetup}. We used an existing symmetric X-ray interferometer \cite{Ferroglio2008} manufactured originally by INRIM to measure the Si lattice parameter. The splitter and mirror lamellas are monolithically connected while the analyser lamella is separate. The lamellas are about 0.73 mm thick and have a spacing of about 10 mm. The analyser is mounted on a piezo-driven tip-tilt platform developed earlier by INRIM \cite{Bergamin2003} and upgraded for the need. It allows to vary both pitch and yaw angle by about $\pm 70\; \mu$rad with sub-nanoradian resolution. A translational piezo stage can vary the axial position of the analyser by a few micrometers with picometer resolution.

The two crystals sit on silicon supports and are held in position by thin films of high-viscosity silicon oil. We arranged them with $\mu$m accuracy using a homemade coordinate measuring machine, which was then removed from the setup. Both crystals have optically polished surfaces on their sides with identical relative orientation to the lattice planes. To pre-align the split crystals, we made these surfaces parallel to each other within one arcsecond using an optical autocollimator. An optical interferometer used the opposite polished surface to monitor the analyser's $\theta$ and $\rho$ angles and the $x$ axial position, allowing for a closed-loop operation.

In the forward exit beam we used a multi channel neutron detector recently developed by the ILL. It is a flat $^3$He detector with 90\% efficiency and 1 mm spatial resolution achieved by crossed wire electrodes. Details of this detector type are given in \cite{buffet2017}.

A vibration isolated bench supports the monochromator crystal, which is located in a thermal neutron beamline, and the optomechanical mounts of the interferometer. To ensure temperature stability, the bench is surrounded by two thermal housings.

An aperture in front of the interferometer reduced the beam size to 1 mm width and 8 mm height, which was the maximum the interferometer could accept (having been designed for X-ray operation). A horizontal slit about one meter upstream controlled the vertical divergence. A slit of 2 mm height left a mean intensity in the two exit beams of 48 counts per second. A 20 mm slit delivered 280 counts per second. We used a neutron wave length of 0.19 nm which corresponds to a Bragg angle of 30$^{\circ}$ on our silicon $\{ 220 \}$ diffracting planes.

\begin{figure*}\centering
\includegraphics[width=0.95\textwidth]{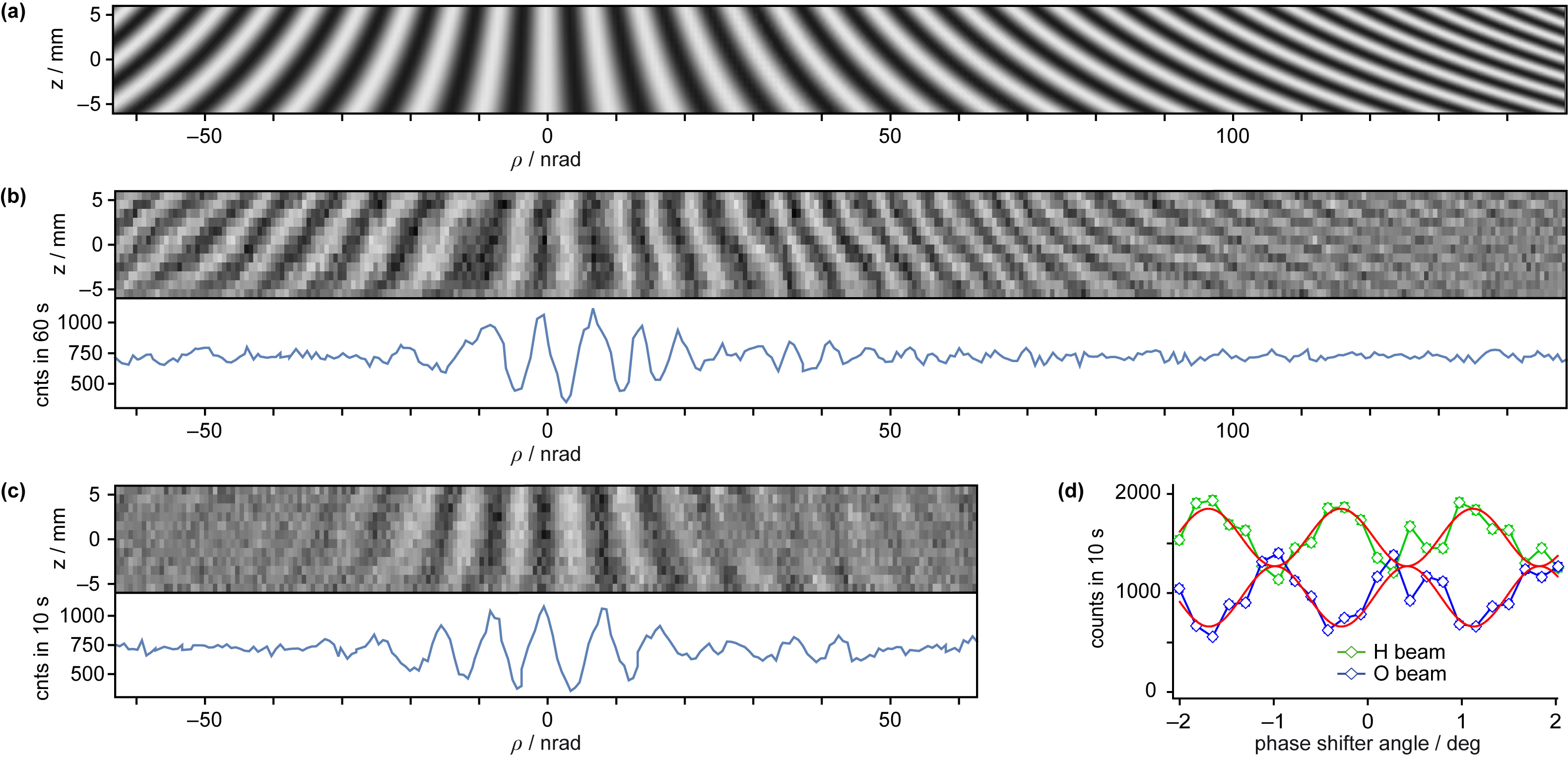}
\caption{Theoretical (a) and observed (b) interference patterns generated during the $\rho$ angle alignment. For each $\rho$ value a neutron-camera image is horizontally integrated yielding a vertical intensity distribution which is shown here by each pixel column. Concatenating the columns of all $\rho$ values generates the complete image (b). The theoretical pattern (a) perfectly reproduces the observed one, although the vertical fringes fluctuate in position and spacing due to phase drifts during the measurement. If the full-beam height is used, the interference can be observed only in the vicinity of perfect $\rho$ alignment (c). The bottom figures in (b) and (c) show the total camera counts. (d) Interferogram created with an auxiliary phase shifter. The red curves are the best sinusoids fitting the data. The fringe visibility is around 40\%.} \label{figresults}
\end{figure*}

\section{Results}

\hspace{1mm}\\
Figure \ref{figresults}(b) shows the interference pattern observed during the alignment of the $\rho$ angle. For this plot, the $\rho$ angle has been scanned in steps of 0.65 nrad and the detector images have been integrated horizontally, leaving a one-pixel column for each $\rho$ value. These columns are plotted next to each other, creating a pattern which originates from two effects.

Firstly, if the $\rho$ angle is misaligned, horizontal moir\'{e} fringes develop
%{\colr along the beam height} Remark: I find this misleading, HL
because the two interfering beams are vertically inclined to each other. As $\rho=0$ rad is approached, the fringe spacing becomes larger and larger until the phase is uniform over the whole beam height. The fringe spacing is given by $\Lambda_z = d/\rho$ if plane waves are assumed. For a divergent beam originating from a point source, the spacing is magnified by $l_D / l_A$ (about 1.55 in our case), where $l_D$ is the distance between source and detector and $l_A$ the distance between source and analyser lamella.

The second effect originates in the fact that the center of the $\rho$ rotation is $z_0 = 38$ mm below the neutron beam. Therefore, each $\rho$ change displaces the analyser along the $x$ axis and
%{\colr Since points located at different heights are displaced by different amounts, the resulting displacement gradient, $\rho(z-z_0)$,} Remark: This sounds unnessesarily complicated to me. HL
causes a phase shift which is visible as  vertical fringes superimposed to the horizontal ones.

Combining both effects we expect the interference pattern cos$[\rho (z\!-\!z_0) 2\pi/d \cdot l_A/l_D]$, which is plotted in Fig.\ \ref{figresults}(a). The observed pattern (b) reproduces the horizontal fringes exactly. The vertical fringes have slightly irregular spacings and positions because the analyser $x$ position and, consequently, the fringe phase were drifting during the five-hour measurement-time.

Figure \ref{figresults}(c) shows the same pattern with the horizontal slit fully open, i.e. with the interferometer illuminated by an incoherent array of point-sources. In this case, the horizontal moir\'{e} pattern, awaited in regions of misaligned $\rho$ angle, disappears. Although each point source creates its own interference pattern, each pattern is vertically shifted and, in the mean, averaged out. Therefore, whenever an extended source (like a neutron source) is used, the interference is only visible in the vicinity of the $\rho=0$ rad alignment. This applies in any case if spatially integrating detectors are used, as shown in the bottom parts of Figs.\ \ref{figresults}(b) and (c).

The motivation of the proof-of-principle experiment was twofold. Firstly, we wanted to understand if the crystal alignment was possible using neutron detection only. For this task we identified an efficient procedure based on the pattern described above. The pre-alignment of the $\rho$ angle, either by using the autocollimator or by measuring neutron intensities, leaves a range of a few $\mu$rad which has to be searched for the 40 nrad wide spot of interference. While scanning the $\rho$ angle, we continuously applied a fast Fourier-transform algorithm to a moving time window of (spatially integrated) neutron counts \cite{Andreas2020}. Being sensitive to periodic modulations of the counts, we could find the correct alignment even if the measurement time per $\rho$ step was so short that the interference amplitude was in the order of the statistical fluctuations. This way we could substantially speed up the $\rho$ scanning.

Secondly, we needed to check the level of the environmental seismic and acoustic noises, as well as their effect on a split-crystal interferometer. The picometer-scale alignment of the two crystals is demonstrated by the neutron interference. In fact, owing to the long neutron count-times, typically 10 s, any high-frequency noise exceeding some tens of picometres would wipe out the interference.

To produce a conventional interferogram, the natural choice would be a controlled $x$ motion of the analyser lamella. Unfortunately, operation in air jeopardized a satisfactory closed-loop operation. In fact, the fluctuating index of refraction limited the stability of the optical interferometer, inducing noise onto the crystal position rather than stabilizing it.

\begin{figure}\centering
\includegraphics[width=0.95\columnwidth]{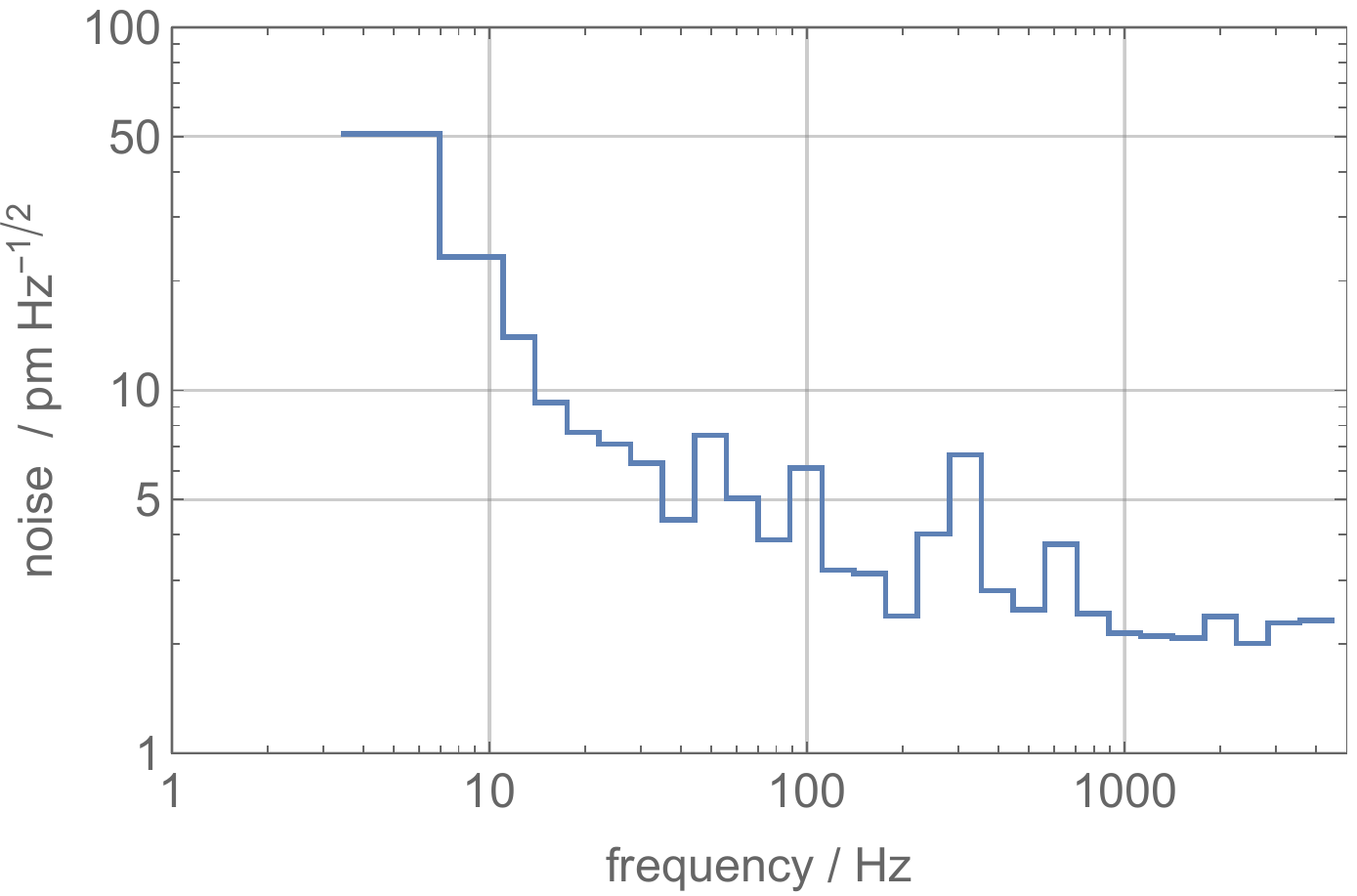}
\caption{Amplitude spectral density of the analyser position-noise. The abscissa is gauged in one-third of octave frequency bands.}\label{fignoise}
\end{figure}

We quantified the noise by looking at the signal of the optical interferometer; the amplitude spectral density of the position noise is shown in Fig.\ \ref{fignoise} over a 4 kHz bandwidth. Expected peaks are observed at 50 Hz and 100 Hz; the broad peak at about 300 Hz might originate in the mechanical resonances of the tip-tilt platform supporting the analyser \cite{Bergamin2003}. The $1/f$ raise of the low-frequency noise originated from both fluctuations of the refractive index of the air and actual mechanical and thermal instabilities of the crystal alignment. Looking at future skew-symmetric realisations, operation in vacuum and optoelectronic feedback of the alignment errors will suppress it.

Eventually, all measurements were made in open-loop mode, and, to scan the interference fringes, we introduced a conventional phase shifter made of a 4 mm thick aluminium slab between second and third lamella. The scan result is shown in Fig.\ \ref{figresults}(d). The expected sinusoidal modulation is disturbed by strong fluctuations that we attribute to low-frequency position-noise of the analyser. Nevertheless, we achieved an interference contrast of around 40\%.

\section{Outlook}

\hspace{1mm}\\
We demonstrated that the alignment of two crystals with the accuracy required for neutron interference is technically possible.
This result is crucial for the development of a split-crystal interferometer having a skew-symmetric geometry and will drive its commissioning. Contrary to the symmetric geometry of our test interferometer, a skew-symmetric design is insensitive to the relative displacements of the split crystals (which can be placed far apart), while displaying the same criticalities regarding their Bragg and pitch alignments. Therefore, long arms are possible and the length can even be varied. Operation in a vacuum will boost the stability and allow a closed-loop operation based on the feedback of the optical-interferometer signals monitoring the Bragg and pitch misalignments. The simultaneous operation of the crystal interferometer with X-rays could monitor systematic errors, e.g., due to non-uniformities of the lattice constant.

We see a large potential to achieve reliable and robust phase measurements even with crystal separations up to the meter scale. Large-scale split-crystal interferometers will open the way to several experiments testing fundamental symmetries and interactions, in particular exploring quantum mechanics and its relation to gravity. The firsts, profiting from the increased area enclosed by the neutron paths, could be more sensitive repetitions of the COW experiment, studying the weak equivalence principle in the quantum regime \cite{saha2014}. Experiments on hypothetical new forces are important tools for probing dark energy fields \cite{lemmel2015,Li2016}. Introducing massive test masses into the interferometer will allow neutron tests of gravitomagnetism \cite{hammad2021}. Our result is a proof of principle showing that such experiments can be realistically considered.

\subsection*{\bf Acknowledgments}

We thank A. Barbone and M. Bertinetti (INRIM) for mechanical manufacturing. The data was taken at the instrument S18 at the ILL and is identified by \cite{S18data}. Open access funding provided by Istituto Nazionale di Ricerca Metrologica within the CRUI-CARE Agreement.

\subsection*{\bf Authors' contributions}

C.\ P.\ Sasso and G.\ Mana developed the theoretical formalism and performed the analytic calculations and numerical simulations. E.\ Massa carried out experimental works that prompted this investigation. H.\ Lemmel and M.\ Jentschel aligned and operated the interferometer, H.\ Lemmel and G.\ Mana simulated the interferometer numerically, C.\ P.\ Sasso and E.\ Massa conceptualized and designed the split-crystal setup and oversaw the assembly and alignment, H.\ Lemmel and E.\ Massa implemented the control software, F.\ Lafont and B. Guerard developed the spatially resolving neutron detector, H.\ Abele, M.\ Jentschel and E.\ Massa organized the collaboration. All authors discussed the results and contributed to the final manuscript.

\bibliography{references}  % Produces the bibliography via BibTeX.

\end{document}